\documentclass[12pt]{article}
\usepackage{amsfonts}
\usepackage{amsmath}
\usepackage{graphicx}

\newtheorem{theorem}{Theorem}

\newtheorem{proposition}[theorem]{Proposition}
\newtheorem{remark}[theorem]{Remark}

\begin{document}

\title{Self-repelling fractional Brownian motion - a generalized Edwards
model for\\
chain polymers}
\author{Jinky Bornales \\
Physics Dept., MSU-IIT, Iligan, The Philippines. \\
E-mail: jinky.bornales@g.msuiit.edu.ph \and Maria Jo\~{a}o Oliveira \\
Univ. Aberta and CMAF, University of Lisbon.\\
E-mail: oliveira@cii.fc.ul.pt \and Ludwig Streit \\
BiBoS, Univ. Bielefeld and CCM, Univ. da Madeira\\
e-mail: streit@uma.pt}
\maketitle

\begin{abstract}
We present an extension of the Edwards model for conformations of individual
chain molecules in solvents in terms of fractional Brownian motion, and
discuss the excluded volume effect on the end-to-end length of such
trajectories or molecules.
\end{abstract}

\section{Introduction}

Individual chain polymers in good solvents are typically modelled by
trajectories of random walks, or \ - in the continuum limit - by Brownian
paths. Such models by themselves however do not take into account that
self-crossings of these paths should be suppressed, this "the excluded
volume" effect will make the trajectories less curly and more extended.
Fractional Brownian paths have been suggested as a heuristic model for such
swelling, or on the other hand for polymers in a collapsed state \cite{Biswas}%
, but a more proper model would be based on self-avoiding random walks, or
on\ a weight factor which penalizes self-crossings, such as in the continuum
Edwards \cite{Bolthausen} \cite{Edwards} \cite{Varadhan} \cite{West1} \cite%
{West3} \cite{westwater} or the discrete Domb-Joyce \cite{Domb} model.

The ensuing swelling of the molecular conformations is given by the Flory
index \cite{Fisher69} \cite{Flory} which describes the scaling of the
end-to-end distance as a function the number of monomers. It has been
extensively studied both in the (chemical) physics and the mathematics
community. The physics literature is characterized by structural intuition
and far-reaching predictions, the mathematical results are less far-reaching
but provide the high reliability characteristic of the mathematical
approach. Both are too vast to be quoted here, we refer for this to recent
reviews \cite{Hofstad} \cite{Pelissetto}.

In the present paper, after a few words on fractional Brownian motion fBm,
we shall see that one can extend to fBm the Edwards model of Brownian paths
with exponentially suppressed self-intersections, $\ $a mathematical
existence proof has been established recently \cite{GOSS}. In the third part
of the paper we generalize some by now classical arguments from the physics
literature to explore what the Flory index might be in the fBm case.

\section{The fBm Edwards Model}

\subsection{Fractional Brownian Motion}

Fractional Brownian motion on $\mathbb{R}^{d}$, $d\geq 1$, with "Hurst
parameter" $H\in \left( 0,1\right) $ is a $d$-dimensional centered Gaussian
process $B^{H}=\{B^{H}(t):t\geq 0\}$ with covariance function 
\begin{equation*}
\mathbb{E}(B_{i}^{H}(t)B_{j}^{H}(s))=\frac{\delta _{ij}}{2}\left(
t^{2H}+s^{2H}-|t-s|^{2H}\right) ,\quad i,j=1,\ldots ,d,\ s,t\geq 0.
\end{equation*}

For $H=1/2$ it is ordinary $d$-dimensional Brownian motion $B$. We refer to
the recent monographs by Biagini et al.~\cite{oks} and by Y. Mishura \cite%
{Mishura}; for self-intersection local times of fBm see Hu and Nualart \cite%
{HN}.

\subsection{The Edwards Model}

Self-repelling Brownian paths for a time interval $0\leq t\leq l$ can be
modelled via a "Gibbs factor" to suppress self-intersections:%
\begin{equation*}
G=\frac{1}{Z}\exp \left( -g\int_{0}^{l}ds\int_{0}^{l}dt\delta \left(
B(s)-B(t)\right) \right) .
\end{equation*}%
Technically one defines this expression as a limit, using%
\begin{equation*}
\delta _{\varepsilon }(x):=\frac{1}{(2\pi \varepsilon )^{d/2}}e^{-\frac{%
|x|^{2}}{2\varepsilon }},\quad \varepsilon >0,
\end{equation*}%
in particular 
\begin{equation*}
Z=\lim_{\varepsilon \rightarrow +0}\mathbb{E}\left( \exp \left(
-g\int_{0}^{l}ds\int_{0}^{l}dt\delta _{\varepsilon }\left( B(s)-B(t)\right)
\right) \right)
\end{equation*}%
if this quantity is well defined; otherwise a renormalization is required,
as, more generally, in Theorem 2.2 below.

Recently, generalizing an argument\ of Varadhan \cite{Varadhan}, this was
extended in \cite{GOSS} to 
\begin{equation*}
G=\frac{1}{Z}\exp \left( -g\int_{0}^{l}ds\int_{0}^{l}dt\delta \left(
B^{H}(s)-B^{H}(t)\right) \right) ,
\end{equation*}%
as follows.

\begin{theorem}
The Edwards model is well defined for all $H<1/d$, with 
\begin{equation*}
G=\frac{1}{Z}\exp \left( -g\int_{0}^{l}ds\int_{0}^{l}dt\delta \left(
B^{H}(s)-B^{H}(t)\right) \right) .
\end{equation*}
\end{theorem}

\begin{theorem}
For $H=1/d$ and $g$ sufficiently small%
\begin{equation*}
G=\lim_{\varepsilon \searrow 0}\frac{1}{Z_{\varepsilon }}\exp \left(
-g\int_{0}^{l}ds\int_{0}^{l}dt\delta _{\varepsilon }\left( B^{H}(s)-B^{H}(t)%
\text{ }\right) \right) \ ,
\end{equation*}%
with 
\begin{equation*}
Z_{\varepsilon }\equiv \mathbb{E}\left( \exp \left(
-g\int_{0}^{l}ds\int_{0}^{l}dt\delta _{\varepsilon }\left( B^{H}(s)-B^{H}(t)%
\text{ }\right) \right) \right)
\end{equation*}%
is well-defined.
\end{theorem}

\section{The Flory Index}

When the number $N$ of monomers of a polymer becomes large one expects its
end-to-end length $R$ to scale \cite{deGennesbook}%
\begin{equation*}
R(N)\sim N^{\upsilon }.
\end{equation*}%
For (fractional) Brownian motion the root-mean-square length 
\begin{equation*}
R=\sqrt{\mathbb{E}\left( B^{H}(N)^{2}\right) }
\end{equation*}%
\ is scaling with%
\begin{equation*}
\upsilon =H.
\end{equation*}

But the excluded volume effect makes the paths and polymers swell: the
end-to-end length increases. For the Brownian motion case there is the
famous Flory formula 
\begin{equation*}
\upsilon =\upsilon \ (d)=\frac{3}{d+2}
\end{equation*}%
based originally on a mean field argument. Since its proposal by Flory \cite%
{Fisher} \cite{Flory}, numerous methods were invoked to put it on a more
solid mathematical basis, a process which has up to now been fully
successful in the case $d=1$ \cite{Hofstad}.

To obtain what may be considered as a first guess of a similar formula for
fBm we shall return to the modest beginnings, generalizing Fisher's original
argument \cite{Fisher} \cite{Fisher69} (see e.g.~the review given in McKenzie 
\cite{McKenzie}) to the case at hand.

\subsection{The Fisher Argument}

A partition function $Z(R)$ for a freely jointed chain of $N$ segments for
which the end-to-end length has fixed modulus $R$ is given by

\begin{equation*}
Z(R)=aR^{d-1}\exp (-\frac{dR^{2}}{2N}),
\end{equation*}%
and leads to a free energy%
\begin{equation*}
\ F_{1}=-\ln Z\sim \frac{dR^{2}}{2N}-(d-1)\ln R.
\end{equation*}%
Instead of such a chain a continuous model is that of a Brownian trajectory
from time zero to time $N$, for which one computes%
\begin{equation}
\mathbb{E}\left( \delta \left( B\left( N\right) -\vec{R}\right) \right)
=\left( 2\pi N\right) ^{-d/2}\exp \left( -\frac{R^{2}}{2N}\right) .
\end{equation}%
For the fBm case this formula generalizes to%
\begin{equation}
\mathbb{E}\left( \delta \left( B^{H}\left( N\right) -\vec{R}\right) \right)
=\left( 2\pi N^{2H}\right) ^{-d/2}\exp \left( -\frac{R^{2}}{2N^{2H}}\right)
\label{f}
\end{equation}%
from which we see that $N\rightarrow N^{2H},$ and hence we should consider%
\begin{equation*}
Z(R)=aR^{d-1}\exp (-\frac{dR^{2}}{2N^{2H}})
\end{equation*}%
i.e.%
\begin{equation*}
F_{1}=-\ln Z\sim \frac{dR^{2}}{2N^{2H}}-(d-1)\ln R.
\end{equation*}%
For the repulsive excluded volume energy of fBm paths $x$ with $x(N)=\vec{R}%
, $%
\begin{equation*}
F_{2}=-\ln \mathbb{E}_{x(N)=\vec{R}}\left( \exp \left(
-g\int_{0}^{N}ds\int_{0}^{N}dt\delta \left( x(s)-x(t)\right) \right) \right)
\end{equation*}%
dimensional considerations and mean field arguments \cite{McKenzie} suggest%
\begin{equation*}
F_{2}\sim const.\frac{N^{2}}{R^{d}}
\end{equation*}%
Maximizing%
\begin{equation*}
F(N,R)=F_{1}(N,R)+F_{2}(N,R)
\end{equation*}%
with regard to $R$ leads to%
\begin{equation*}
0=\frac{dR}{N^{2H}}-\frac{d-1}{R}-const.N^{2}R^{-d-1}.
\end{equation*}%
Assuming that the 2nd term is negligible one finds%
\begin{equation*}
R^{d+2}\sim N^{2H+2}
\end{equation*}%
i.e.%
\begin{equation*}
R\sim N^{\upsilon _{H}}
\end{equation*}%
with%
\begin{equation}
\upsilon _{H}(d)=\frac{2H+2}{d+2}.  \label{(1)}
\end{equation}

\begin{remark}
A polymer model with $B(t^{2H})$ instead of $B^{H}(t)$ would produce the
same expression as in (\ref{f}), hence also the same Flory index, but would
not share the homogeneity implied by the stationary increments of $%
B^{H}(\cdot ).$
\end{remark}

\subsection{The Critical Dimension}

The derivation of $\upsilon _{H}$ is evidently heuristic and needs
validation. For this it is worth noting that for Brownian motion there is a
critical dimension $d_{c}=4$ defined by the fact that for $d\geq d_{c}$
there is no excluded volume effect, so that $R$ scales like the unperturbed
Brownian motion: 
\begin{equation*}
R\sim N^{1/2}
\end{equation*}%
i.e. 
\begin{equation*}
\upsilon _{1/2}(4)=1/2.
\end{equation*}%
We can ask for which dimension, more generally, the fBm Flory index will
show no excluded volume effect from self-crossings, i.e. 
\begin{equation*}
\upsilon _{H}(d_{c})=H.
\end{equation*}%
Inserting our ansatz (\ref{(1)}) one finds 
\begin{equation*}
Hd_{c}=2.
\end{equation*}%
and indeed it is known (Theorem 1.1 of Talagrand \cite{Talagrand}) that $d$%
-dimensional fBm has no double points iff%
\begin{equation*}
Hd\geq 2,
\end{equation*}%
in other words, our $\upsilon _{H}$ predicts $d_{c}$ correctly.

\begin{remark}
As a consequence, any Flory formula should be considered only up to the
critical dimension, i.e. as long as there are double points and an excluded
volume effect. Similarly, any prediction of $\upsilon _{H}>1$ would be
unphysical: the end-to-end distance cannot grow faster than the number N of
monomers. In the case at hand this suggests for the one-dimensional case%
\begin{equation*}
\upsilon _{H}(1)=\left\{ 
\begin{array}{cc}
\frac{2H+2}{3} & \text{ if }H\leq \frac{1}{2} \\ 
1 & \text{if }H>\frac{1}{2}%
\end{array}%
\right. .
\end{equation*}%
Note that for small $H$ the scaling exponent as predicted would be strictly
less than one while for the Brownian motion case 
\begin{equation*}
\upsilon _{1/2}(1)=1
\end{equation*}%
has been proven \cite{Hofstad} \cite{westwater}. The infimum%
\begin{equation*}
\lim_{H\rightarrow 0}\upsilon _{H}(1)=2/3
\end{equation*}%
happens to be the scaling exponent of the myopic random walk \cite{Hofstad}.
\end{remark}

\begin{remark}
In the attached figure 1 the two red lines correspond to $%
\upsilon _{H}(d)=1$ and to the critical dimension as a function of the Hurst
index $H$, respectively. Above these the Flory index is unphysical. On the
green lines $\upsilon _{H}$ is validated. (The existence proof of the fBm
Edwards model in \cite{GOSS} works below the dashed line.)
\end{remark}

\begin{remark}
For fixed dimension $d$, any extension $F(H)$ of the Flory formula to
general Hurst indices $H$ will have to obey%
\begin{equation}
F(\frac{1}{2})=\frac{3}{d+2}  \label{FF}
\end{equation}%
for the usual Brownian motion (Flory-Fisher), and 
\begin{equation}
F(\frac{2}{d})=\frac{2}{d}  \label{cp}
\end{equation}%
at the critical point (Talagrand). Note that our ansatz (\ref{(1)}) 
\begin{equation*}
F(H)=\upsilon _{H}(d)=\frac{2H+2}{d+2}
\end{equation*}%
is just the unique linear interpolation between those two values.
\end{remark}

\subsection{A Recursion Formula}

For $H=1/2$ Kosmas and Freed \cite{Kosmas} derive a recursion formula 
\begin{equation}
2-\frac{1}{\upsilon (d)}=\frac{4-d}{3}\left( 2-\frac{1}{\upsilon \left(
1\right) }\right)  \tag{(4.13)}  \label{(4.13)}
\end{equation}%
(Here and in the following we label formulas from - or analogous to those in
- the paper \cite{Kosmas} by their numbers in that article, in double
brackets.)

The derivation of this formula is specific to the Brownian motion case and
does not hold for general $\upsilon _{H}$. Hence in what follows we shall
generalize their arguments which led to (\ref{(4.13)}) to first obtain a
valid recursion formula and then check whether it is satisfied by $\upsilon
_{H}$ \ as given in (\ref{(1)}).

We begin by considering%
\begin{equation*}
Z\left( g,N\right) \equiv \mathbb{E}\left( \exp \left(
-g\int_{0}^{N}ds\int_{0}^{N}dt\delta \left( B^{H}(s)-B^{H}(t)\right) \right)
\right) .
\end{equation*}%
From the defining relation%
\begin{equation*}
\mathbb{E}\left( B^{H}(s)B^{H}(t)\right) =\frac{1}{2}\left(
s^{2H}+t^{2H}-\left\vert s-t\right\vert ^{2H}\right)
\end{equation*}%
we see that for $a>0$ the processes $\left\{ B^{H}(t):t>0\right\} $ \ and $%
\left\{ a^{-H}B^{H}(at):t>0\right\} $ obey the same law. Making this
substitution and a change of integration variables $as=\sigma ,$ $at=\tau $
we obtain%
\begin{eqnarray*}
Z\left( g,N\right) &=&\mathbb{E}\left( \exp \left(
-ga^{Hd-2}\int_{0}^{aN}ds\int_{0}^{aN}dt\delta \left(
B^{H}(s)-B^{H}(t)\right) \right) \right) \\
&=&Z\left( a^{Hd-2}g,aN\right) .\qquad\qquad\qquad\qquad\qquad\qquad\qquad\qquad ((2.14))
\end{eqnarray*}

Likewise we find for the mean-square end-to-end distance 
\begin{equation}
\left\langle R^{2}\right\rangle \equiv \frac{1}{Z\left( g,N\right) }\mathbb{E%
}\left( \left( B^{H}(N)\right) ^{2}\exp \left(
-g\int_{0}^{N}ds\int_{0}^{N}dt\delta \left( B^{H}(s)-B^{H}(t)\right) \right)
\right)  \label{(2)}
\end{equation}

\begin{eqnarray*}
\left\langle R^{2}\right\rangle &=&a^{-2H}\mathbb{E}\left( \left(
B^{H}(aN)\right) ^{2}\exp \left(
-ga^{Hd-2}\int_{0}^{aN}ds\int_{0}^{aN}dt\delta \left(
B^{H}(s)-B^{H}(t)\right) \right) \right)  \notag \\
&=&a^{-2H}f\left( a^{Hd-2}g,aN\right)  \qquad\qquad\qquad\qquad\qquad\qquad\qquad\qquad((2.18)) \\
&=&N^{2H}f\left( N^{2-Hd}g,1\right) .  \qquad\qquad\qquad\qquad\qquad\qquad\qquad\qquad\ ((2.19))
\end{eqnarray*}

(Note the critical dimension $Hd=2\ $where $\left\langle R^{2}\right\rangle
^{1/2}\sim N^{H}$.) For large $N$ one expects a power law behavior for the
unknown function $f$, i.e.%
\begin{equation}
\left\langle R^{2}\right\rangle \sim N^{2H}\left( N^{2-Hd}g\right) ^{x} 
\tag{(2.22)}  \label{(2.22)}
\end{equation}%
with an exponent $x$ to be determined.

As a next step we restrict one coordinate of the positions, $%
x_{i}(t)=B_{i}^{H}(t)$ \ to the interval $[0,D]$ \ by inserting 
\begin{equation*}
1_{[0,D]}(B_{i}^{H})=\left\{ 
\begin{array}{cc}
1 & \text{if }B_{i}^{H}(t)\in \lbrack 0,D]\text{ for all }t \\ 
&  \\ 
0 & \text{ otherwise}%
\end{array}%
\right.
\end{equation*}%
into (\ref{(2)}). One obtains%
\begin{gather}
\left\langle R^{2}\right\rangle _{D}=\frac{1}{Z_{D}\left( g,N\right) }%
\mathbb{E}\left( 1_{[0,D]}(B_{i}^{H})\left( B^{H}(N)\right) ^{2}\exp \left(
-g\int_{0}^{N}ds\int_{0}^{N}dt\delta \left( B^{H}(s)-B^{H}(t)\right) \right)
\right)  \notag \\
=\frac{a^{-2H}}{Z_{D}\left( g,N\right) }\mathbb{E}\left(
1_{[0,a^{H}D]}(B_{i}^{H})\left( B^{H}(aN)\right) ^{2}\exp \left( -a^{Hd-2}%
\underset{}{g\int_{0}^{aN}}d\sigma \int_{0}^{aN}d\tau \delta \left(
B^{H}(\sigma )-B^{H}(\tau )\right) \right) \right)  \notag \\
=a^{-2H}F(a^{H}D,\ a^{Hd-2}g,aN)=N^{2H}F(N^{-H}D,\ N^{2-Hd}g,1).  \tag{(4.3)}
\label{(4.3)}
\end{gather}

Now assume that asymptotically there is a dimensionless correction factor $h$
for 
\begin{equation*}
\left\langle R^{2}\right\rangle _{D}\approx \left\langle R^{2}\right\rangle
h\left( \frac{D}{\sqrt{\left\langle R^{2}\right\rangle }}\right) .
\end{equation*}%
It should grow as $D$ becomes small which suggests a power law behavior for
the function $h$:
\begin{equation}
\left\langle R^{2}\right\rangle _{D}\approx \left\langle R^{2}\right\rangle
\left( \frac{D}{\sqrt{\left\langle R^{2}\right\rangle }}\right) ^{-y} 
\tag{(4.5)}  \label{(4.5)}
\end{equation}%
with $y$ to be determined. As $D$ \ approaches a minimal value $D_{0}$ -
approximately the extension of a monomer ("Kuhn length") - the polymer
becomes effectively $\left( d-1\right) $-dimensional:

\begin{equation}
\left\langle R_{d-1}^{2}\right\rangle \approx \left\langle
R_{d}^{2}\right\rangle _{D}\approx D_{0}^{-y}\left\langle
R_{d}^{2}\right\rangle ^{1+\frac{y}{2}}.  \tag{(4.6)}  \label{(4.6)}
\end{equation}%
This provides a relation between the end-to-end length for dimensions $d$
and $d-1$. To obtain from this a recursion relation, recall equation (\ref%
{(2.22)}):%
\begin{equation*}
\left\langle R_{d}^{2}\right\rangle \sim N^{2H+x(2-Hd)}g^{x}
\end{equation*}%
and introduce\ instead of $x$ \ the (unknown)%
\begin{equation*}
2\upsilon _{H}(d)\equiv 2H+x(2-Hd)
\end{equation*}%
so that 
\begin{equation}
\left\langle R_{d}^{2}\right\rangle =c_{d}N^{2\upsilon _{H}(d)}g^{\frac{%
2\upsilon _{H}(d)-2H}{2-Hd}}  \tag{(4.8)}  \label{(4.8)}
\end{equation}%
and$\ $%
\begin{equation}
\left\langle R_{d-1}^{2}\right\rangle =c_{d-1}N^{2\upsilon _{H}(d-1)}g^{%
\frac{2\upsilon _{H}(d-1)-2H}{2-H\cdot \left( d-1\right) }}.  \tag{(4.9)}
\label{(4.9)}
\end{equation}%
On the other hand from (\ref{(4.6)}) we have%
\begin{equation}
\left\langle R_{d-1}^{2}\right\rangle \approx D_{0}^{-y}\left\langle
R_{d}^{2}\right\rangle ^{1+\frac{y}{2}}=const.N^{2\upsilon _{H}(d)\left( 1+%
\frac{y}{2}\right) }g^{\frac{2\upsilon _{H}(d)-2H}{2-Hd}\left( 1+\frac{y}{2}%
\right) }.
\end{equation}%
Comparing exponents in these two expressions we find 
\begin{eqnarray*}
\upsilon _{H}(d-1) &=&\upsilon _{H}(d)\left( 1+\frac{y}{2}\right) \\
\frac{\upsilon _{H}(d-1)-H}{2-H\cdot \left( d-1\right) } &=&\frac{\upsilon
_{H}(d)-H}{2-Hd}\left( 1+\frac{y}{2}\right) .
\end{eqnarray*}%
The first of these equations gives%
\begin{equation*}
1+\frac{y}{2}=\frac{\upsilon _{H}(d-1)}{\upsilon _{H}(d)},
\end{equation*}%
with this the second one becomes%
\begin{equation*}
\frac{1}{\upsilon _{H}(d-1)}\frac{\upsilon _{H}(d-1)-H}{2-H\cdot \left(
d-1\right) }=\frac{1}{\upsilon _{H}(d)}\frac{\upsilon _{H}(d)-H}{2-Hd}
\end{equation*}%
i.e. this expression does not depend on the dimension $d$ so that all the $%
\upsilon _{H}(d)$ are given in terms of e.g. $\upsilon _{H}(1):$ 
\begin{equation*}
\frac{1}{\upsilon _{H}(d)}\frac{\upsilon _{H}(d)-H}{2-Hd}=\frac{1}{\upsilon
_{H}(1)}\frac{\upsilon _{H}(1)-H}{2-H},
\end{equation*}%
\begin{equation}
\Longrightarrow \upsilon _{H}(d)=\frac{\left( 2-H\right) \upsilon _{H}(1)}{%
\left( d-1\right) \upsilon _{H}(1)+2-dH}.  \label{r}
\end{equation}

\begin{proposition}
\begin{equation*}
\upsilon _{H}(d)=\frac{2H+2}{d+2}
\end{equation*}%
satisfies this recursion equation, with 
\begin{equation*}
\frac{1}{\upsilon _{H}(d)}\frac{\upsilon _{H}(d)-H}{2-Hd}=\frac{1}{2H+2}.
\end{equation*}%
$.$
\end{proposition}

\begin{remark}
The standard Flory index (\ref{(1)}) for $H=1/2$ obeys the recursion formula.
\end{remark}

\begin{remark}
The recursion formula (\ref{r}) implies the correct critical behavior, i.e.
any solution will obey $\upsilon _{H}(d)=H$ for $d=2/H\equiv d_{c}$,
whatever the choice of $\upsilon _{H}(1)$. To see this explicitly, insert $%
d=2/H$ and find
\begin{equation*}
\upsilon _{H}(\frac{2}{H})=\frac{\left( 2-H\right) \upsilon _{H}(1)}{\left(
2/H-1\right) \upsilon _{H}(1)+0}=H.
\end{equation*}
\end{remark}

\begin{remark}
If $\upsilon _{H}(1)$ turned out to be equal to one for all $H$, the
recursion formula would suggest%
\begin{equation*}
\upsilon _{H}(d)=\frac{2-H}{d+1-dH},
\end{equation*}
an expression which then also produces the standard Flory formula for $%
H=1/2, $ as well as the critical dimension $d=2/H.$
\end{remark}

\section{Summary}

The Edwards type model for self-repelling fBm now at hand will raise the
question of how the end-to-end length of trajectories scales as a function
of time (or "number of monomers"). The original Fisher argument, while
criticized regarding its assumptions \cite{Moore} \cite{Cloizeaux}, provides
a simple heuristic "derivation" of the Flory formula which allows an
extension to fBm. The obtained scaling law needs further verification; we
note that it correctly predicts the critical dimension for which the
excluded volume become negligible and obeys a recursion formula based on
dimension reduction. The latter would provide a useful constraint on any
alternate scaling laws.

\begin{center}
\bigskip * * *
\end{center}

J.~B.~and L.~S.~would like to express their gratitude for the kind
hospitality of the Centro de Matem\'{a}tica e Aplica\c{c}\~{o}es
Fundamentais CMAF in Lisbon which made our collaboration possible and
pleasant. This work is supported by PTDC/MAT/100983/2008, ISFL-1-209.

\begin{figure}[h]
\centering
\includegraphics[scale=0.9]{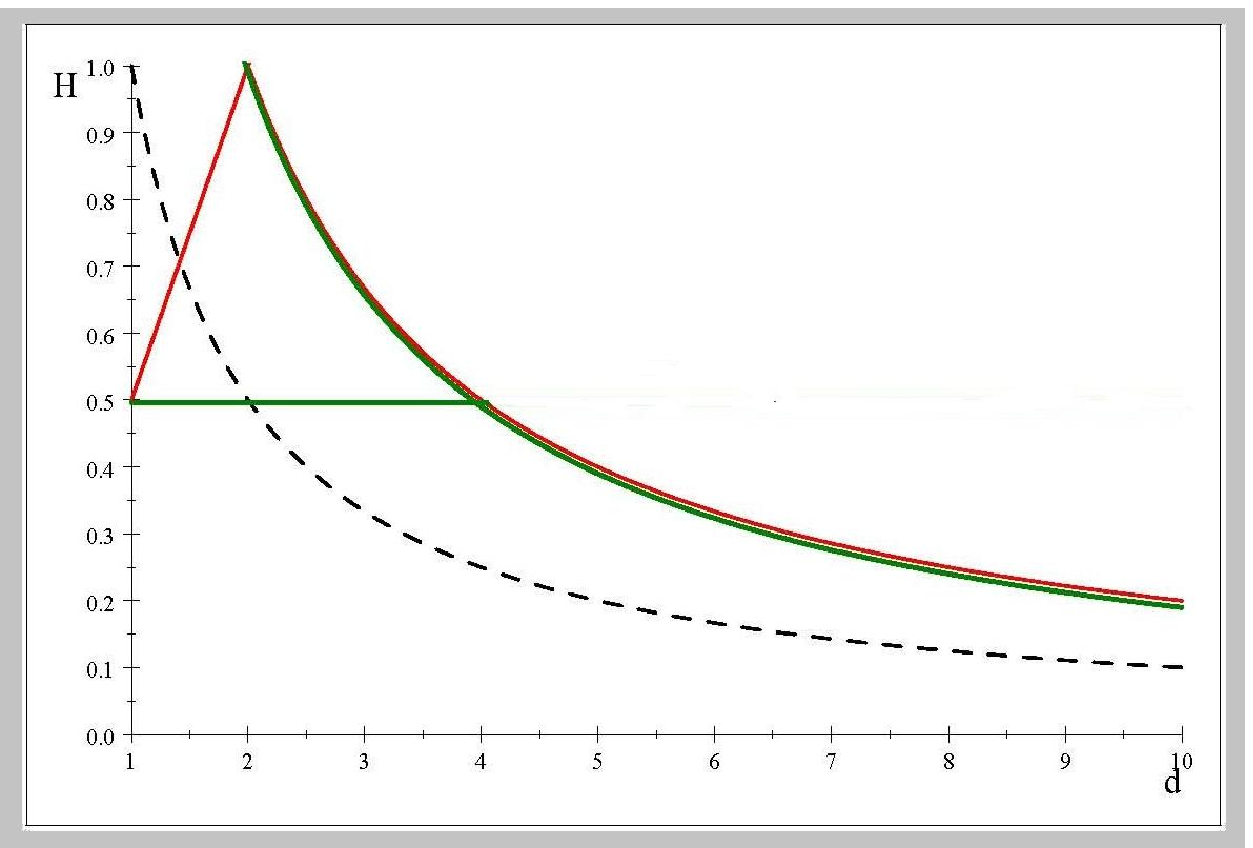}
\caption{The domain of the Flory
index}
\end{figure}


\end{document}